\documentclass[%
prb,
twocolumn,
superscriptaddress,
showpacs,
amsmath,
amssymb,
aps,
floatfix,
]{revtex4-1}

\usepackage{graphicx}
\usepackage{dcolumn} 
\usepackage{bm}      
\usepackage{subfigure}
\usepackage{amsfonts}
\usepackage[usenames]{color}
\usepackage{rotating}
\usepackage{fontenc}
\usepackage{cancel}
\usepackage{amsthm}
\usepackage{hyperref}
\usepackage[normalem]{ulem}
\usepackage{color}
\usepackage{physics}
\usepackage{amsmath}

\definecolor{darkred}{RGB}{139,0,0}
\definecolor{chartreuse}{RGB}{127,255,0}
\definecolor{goldenrod}{RGB}{218,165,32}
\definecolor{gray}{RGB}{127,127,127}

\definecolor{Magenta}{RGB}{255, 0,255}
\definecolor{Orange}{RGB}{255,165, 0}
\definecolor{Gray}{RGB}{127,127,127}

\newcommand{\be}{\begin{equation}}
\newcommand{\ee}{\end{equation}}
\newcommand{\bea}{\begin{eqnarray}}
\newcommand{\eea}{\end{eqnarray}}
\newcommand{\bw}{\begin{widetext}}
\newcommand{\ew}{\end{widetext}}

\newcommand{\bi}{\begin{itemize}}
\newcommand{\ei}{\end{itemize}}

\begin{document}

\title{Fractal Spectrum of the Aubry-Andr\'{e} Model}

\author{Ang-Kun Wu}
\affiliation{Department of Physics and Astronomy, Center for Materials Theory, Rutgers University, Piscataway, NJ 08854 USA}

\date{\today}

\begin{abstract} 
The Aubry-Andr\'{e} model is a one-dimensional lattice model for quasicrystals with localized and delocalized phases. At the localization transition point, the system displays fractal spectrum, which relates to the Hofstadter butterfly. In this work, we uncover the exact self-similarity structures in the energy spectrum. We separate the fractal structures into two parts: the fractal filling positions of gaps and the scaling of gap sizes. We show that the fractal fillings emerge for a certain type of incommensurate periodicity regardless of potential strength. However, the power-law scaling of gap sizes is characteristic for general incommensurability at the critical point of the model.
\end{abstract}

\maketitle


Physical systems with incommensurability demonstrate novel physical properties, e.g. the bilayer systems with incommensurate Moir\'{e} superlattices \cite{PhysRevB.103.155157,PhysRevResearch.2.023325,fu2020magic},topological materials driven by quasiperiodic potential \cite{PhysRevB.104.L041106,PhysRevB.101.235121,PhysRevLett.120.207604} and strongly correlated quasicrystals \cite{PhysRevLett.126.040603} with non-Fermi-liquid behavior \cite{PhysRevB.92.224409,deguchi2012quantum}. In the exploration of correlation effects in one-dimensional quasicrystals, we found that the incommensurability between the discrete lattice and the onsite potential gives rise to self-similar energy spectrum, which creates singularities and changes correlation fundamentally \cite{PhysRevB.100.165116}. It turns out that the filling position of conduction electrons is critical for the study of correlations. A casual choice of the electron filling will gap the system by an emerging mini-gap from increasing system sizes. Besides, in the quest of the impurity problem on such one-dimensional quasicrystals, the numerical renormalization group asks for the spectrum logarithmically close to the Fermi energy, which is challenging with all the structures emerging from higher spectrum resolutions. Thus, the understanding of the fractal spectrum in the thermodynamic limit becomes important for the strongly correlated quasicrystals.

The study of the fractal spectrum could be traced back to the incommensurate magnetic fields upon two-dimensional Bloch electrons, known as the Hofstadter butterfly\cite{PhysRevB.14.2239}. In the problem, the Schr\"{o}dinger equation turns into the Harper's equation \cite{PhysRevB.13.1595,harper1955single,Azbel1964,rauh1974degeneracy}, a one-dimensional difference equation. By varying the magnetic field, the periodicity of the equation is continuously changed and the Hofstadter butterfly could be generated by the stacking the energy spectrum for each field.  Hofstadter, as well as others\cite{Azbel1964,xu1986fractal}, summarized rules based on empirical examination of the numerical data to construct the butterfly. From the ``skeleton" of the butterfly, finer structures could be created by two transformation rules, which give the spectrum by the L(left) and R(right) type or the C(center) type of the ``skeleton". These rules, in principle, could generate the butterfly and the spectrum of a particular field indefinitely, but they requires the knowledge of the spectrum of child field parameters for a parent field and lack the knowledge of electron fillings. The Harper's difference equation is strikingly similar to the Aubry-Andr\'{e} (AA) model at the localization critical point and sometimes are also called Aubry-Andr\'{e}-Harper model.
The incommensurate magnetic field problem is also studied on layered systems with Moir\'{e} pattern\cite{dean2013hofstadter,PhysRevB.84.035440,lu2021multiple}. 

In this work, we aim for providing self-contained rules to reconstruct the fractal spectrum of the one-dimensional quasicrystal, the AA model \cite{PhysRevLett.43.1954,aubry1980analyticity,roati2008anderson}, to the thermodynamic limit. Instead of viewing the fractal spectrum empirically, we associated the rational periodicity with the number of electrons in a unit cell and the irrational periodicity will generate the mini-bands pattern scaling with the system size.
We argue that the self-similar spectrum exists only for a particular type of incommensurability (irrational periodicity). The fractal filling of electrons depends only on the irrational periodicity and exists beyond the critical point. Morerover, the power-law decay of gap sizes is characteristic at the critical point of the AA model and makes the spectrum fractal for proper incommensurability.

This paper is organized as follow. First, we introduce the Aubry-Andr\'{e} model and the multi-band theory for rational periodicity. The organizing rules of the spectrum are summarized for arbitrary periodicities.
Second, we show the fractal spectrum for a typical irrational periodicity and derive the self-similarity in the filling positions of gaps from the organizing rules. We argue that the self-similar spectrum is limited to a certain type of irrational periodicities.
Third, we identify the scaling behavior of the sizes of the emerging mini-gaps across different phases of the AA model and show that the power-law scaling is characteristic at the critical point. Conclusions and discussions will be given in the end.


\textit{The model and organizing rules of bands.}
A general one-dimensional tight-bind model with a constant hopping parameter and periodic onsite potential could be written as
\begin{equation}
    \begin{split}
        H=-t\sum_{j}(c_{j}^\dagger c_{j+1}+h.c.)+\sum_j f(j)c_j^
        \dagger c_j,
    \end{split}
\end{equation}
where $f(j)$ is some periodic function with rational or irrational frequency $\alpha$ so that $f(j+1/\alpha)=f(j)$. A typical function is the cosine function, 
\begin{equation}
    \begin{split}
        f(j)=\lambda \cos(2\pi \alpha j+\phi),
    \end{split}
\end{equation}
known as the Aubry-Andr\'{e} model, where $\alpha$ is an irrational number for quasicrystals, i.e. $\alpha=(\sqrt{5}-1)/2$, and $\phi$ is a random phase that does not change the energy spectrum fundamentally. Without particular mention, the default $\phi$ is set to zero. $\lambda$ is the strength of the quasi-periodic potential relative to the hopping ($t=1$ by default). The Fourier transformation of the AA model reveals the self-duality of the model \cite{aubry1980analyticity} between the momentum and real space, where $\lambda$ is mapped to $2t$ in the momentum space. This self-duality identifies the localization transition point at $\lambda=2t$. For $\lambda<2t$,  all the wavefunctions are localized in the momentum space while for $\lambda>2t$, all the wavefunctions are localized in the real space. 

Before we start to dive into the irrational part of the AA model, it is relatively easy to understand this model with rational frequencies, $\alpha=p/q$ with $p,q$ are coprime integers. The simplest example is $\alpha=1/2$ and the onsite potentials oscillates between $-\lambda, \lambda$ along the chain. We could solve this model analytically by considering the unit cell of length $2d$ (assuming the nearest neighbor distance is $d$) and each unit cell has two atoms. If we denote the annihilation operators for the two atoms as $a_j,b_j$, 
\begin{equation}
        H=-t\sum_j(a_j^\dagger b_j+ b_j^\dagger a_{j+1}+h.c.)+\lambda[\sum_j a_j^\dagger a_j - b_j^\dagger b_j],
\end{equation}
whose continuum model could be written as bilinear $[a_k,b_k]'$ with $k$ as the momentum indices
\begin{equation}
    \begin{split}
        H(k)&=\begin{pmatrix}
        \lambda & -2t\cos(kd)\\
        -2t\cos(kd) & -\lambda
        \end{pmatrix}
    \end{split}.
\end{equation}
It is easy to see that there are two bands from the continuum spectrum $E_{\pm} = \pm\sqrt{\lambda^2+4t^2\cos^2(kd)},$ with $k=\frac{2\pi n}{Ld},kd\in [\frac{2\pi}{L},2\pi].$ The gap appears at a half filling with size $\Delta=2\lambda$. 
For an arbitrary rational frequency $\alpha=p/q$, we have corresponding $q-$band continuum in the thermodynamic limit:
\begin{equation}
    \begin{split}
        H(k)=\begin{pmatrix}
        V_1 & -te^{ikd} & 0 & \cdots & -te^{-ikd}\\
         -te^{-ikd} & V_2 & -te^{ikd} & 0 & \cdots \\
         \vdots & \vdots & \vdots & \vdots & \vdots\\
         -te^{ikd}& 0 &\cdots & -te^{-ikd} & V_q
        \end{pmatrix}
    \end{split},
\end{equation}
where $V_j=\lambda\cos(2\pi \frac{p}{q}j+\phi), i=1,2,\cdots,q$. For some periodicity with a certain phase, e.g. $\alpha=1/4,\phi=0$, there are bands merging due to the same potentials at $j=1+4n,3+4n$. However, this symmetry can be broken by changing the phase $\phi$. 

From the above unit cell viewpoint, rules could be summarized for rational frequencies: (i) For $\alpha=p/q$, the periodicity of the potential is $q$ and it should generate $q$ bands for a general phase $\phi$. Namely, electrons are filled equally into $q$ bands with gaps appearing at fillings $1/p,2/p,\cdots,(p-1)/p$. (ii) For arbitrary $\phi$, the models between $\alpha, 1-\alpha$ differ by a phase shift of $2\phi$. Namely, the spectrums of $\alpha$ and $1-\alpha$ are symmetric. 
Combining the above two rules, the gap filling positions could be summarized for an arbitrary $\alpha$ whose fractional representation is unknown, or irrational $\alpha$:
(iii) The spectrum is separated by bands of fillings $\{\alpha\},\{2\alpha\},\cdots;\{1-\alpha\},\{2(1-\alpha)\}\cdots$ with $\{\cdots\}$ denotes the fractional part of the number.

\begin{figure}[t!]
\begin{center}
\includegraphics[width = 0.45\textwidth]{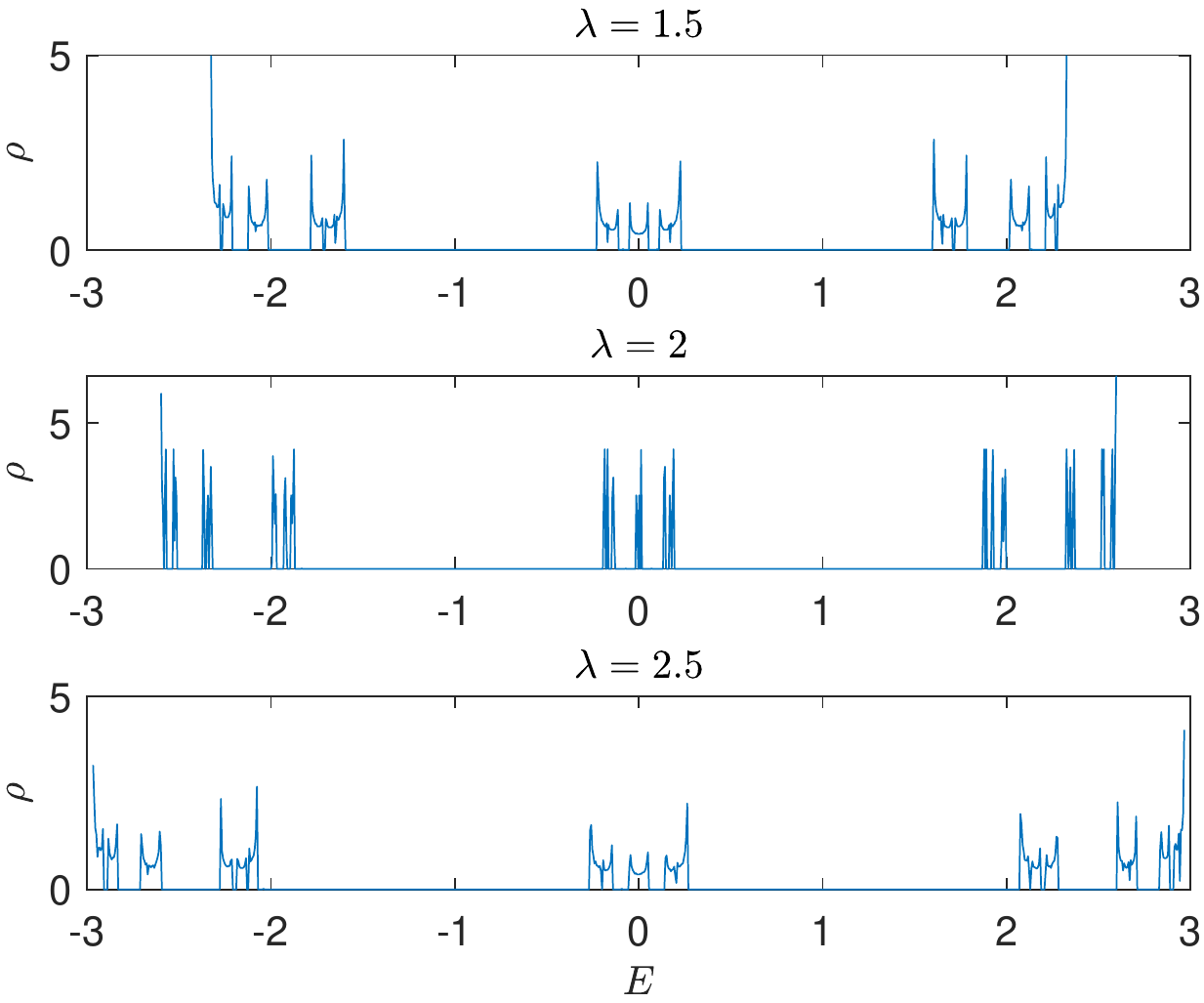}
\includegraphics[width = 0.45\textwidth]{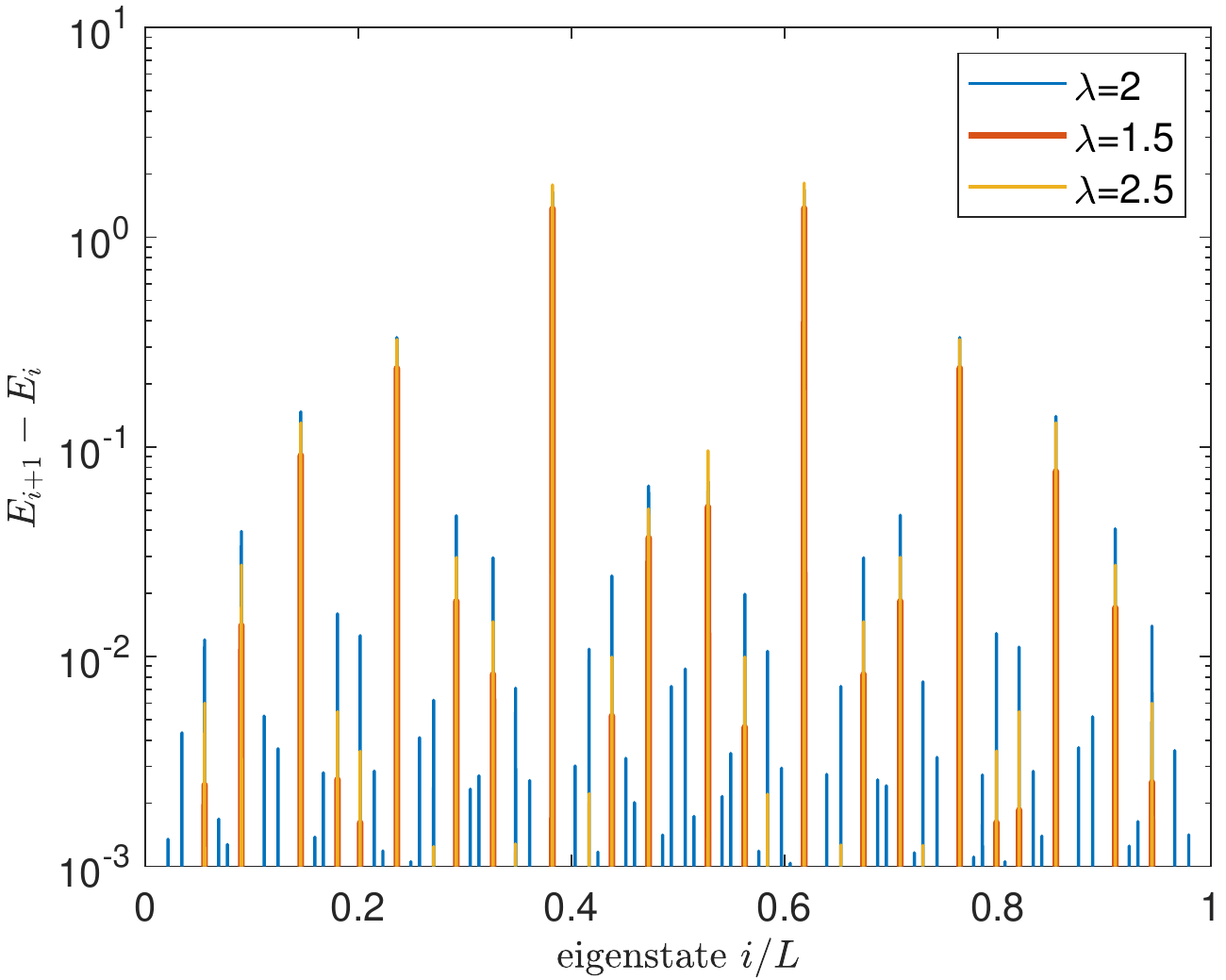}
\caption{The DOS and nearest state gaps of the AA model at different phases with system size $L=10,000,\alpha=(\sqrt{5}-1)/2$ and $\phi=0$. Upper panel: the density of states (DOS). Lower panel:the nearest state gaps with peaks locating around $i=[N\{\alpha\}],[N\{2\alpha\}],\cdots;[N\{1-\alpha\}],[N\{2(1-\alpha)\}],\cdots$, where $[\cdots]$ denotes the largest integer smaller the number. }
\label{lamsDemo}
\end{center}
\end{figure}

The rule (iii) is the most important rule for our later discussion about the gap size scaling. Empirically, the sizes of gaps appear in a descending manner by the above order. For instance, the largest gaps start at fillings $\{\alpha\},\{1-\alpha\}$, and smaller gaps at $\{2\alpha\},\{2(1-\alpha)\}$ and so on. 
By definition, for an irrational $\alpha$, the two filling series will continues indefinitely and give infinite number of gaps in the thermodynamic limit, regardless of the quasi-periodic strength $\lambda$. As shown in Fig. \ref{lamsDemo}, the filling of the first several sharp gaps coincide at the same position for different $\lambda$ with the same frequency. However, the densities of states (DOS) are quite different between the critical point ($\lambda=2$) and the other two phases. Note that the sizes of the nearest state gaps are quite different for different phases--the gap sizes decay in a slower fashion at the critical point.

\begin{figure}[t!]
\begin{center}
\includegraphics[width = 0.49\textwidth]{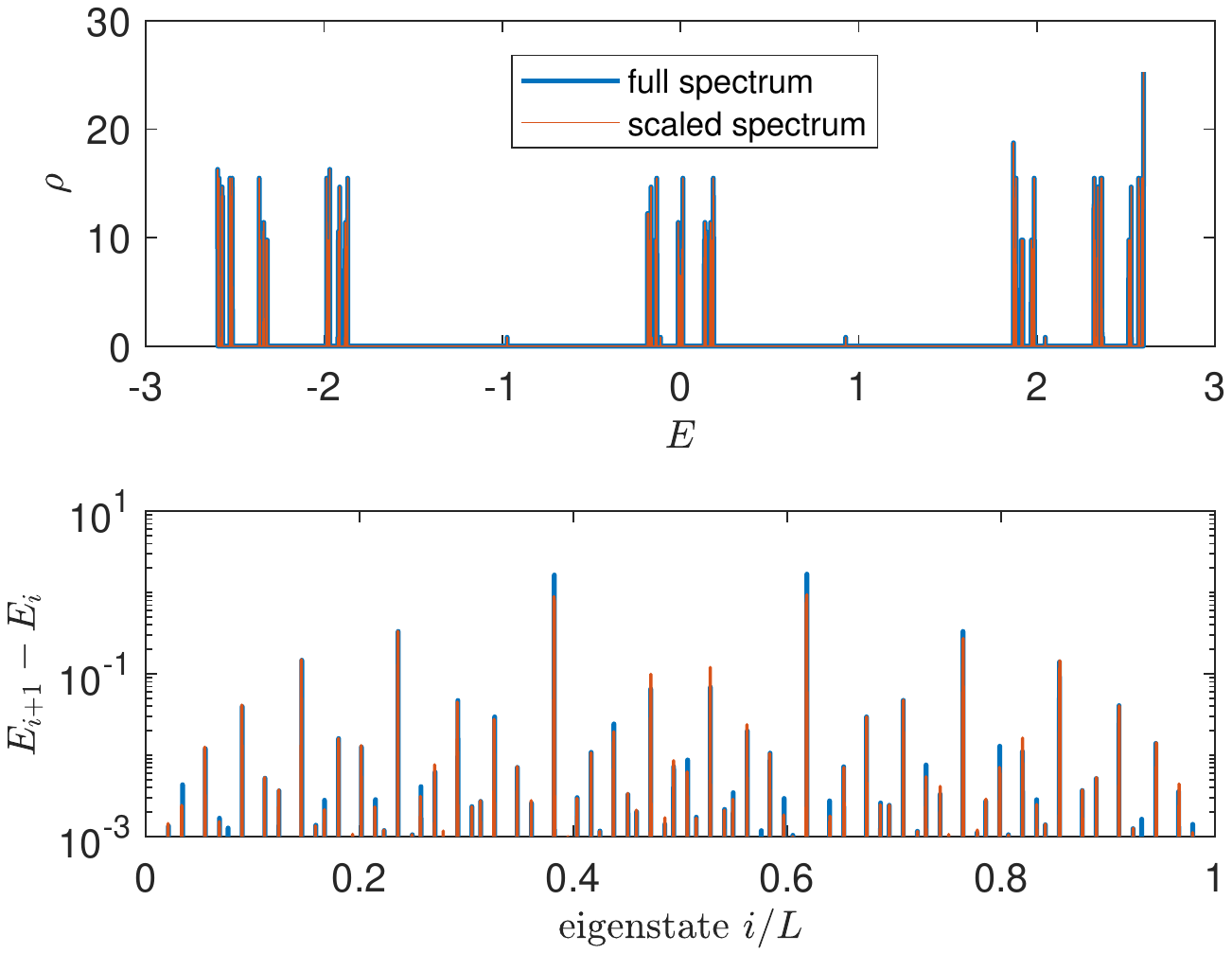}
\caption{The full fractal spectrum at the critical point with system size $L=10,000,\alpha=(\sqrt{5}-1)/2$ and averaged over $100$ random phases. Upper panel: the density of states (DOS) of the full spectrum and the scaled center spectrum (filling=$[0.3821,0.6181]$ with edge energy $\pm 0.1891$ and scaled by factor $13.74=2.5975/0.1891$). Lower panel:the nearest state gaps of the full and scaled spectrum.}
\label{fullspec}
\end{center}
\end{figure}

The self-similarity of the AA model at the critical point is indicated in Fig. \ref{fullspec}. If we zoom in the center part of the full spectrum and rescale the center part to the whole band size (red curves), there is a clear reconstruction of the whole spectrum from the middle part in terms of both the DOS and the nearest state gaps. Mismatches of the gap sizes are due to the finite lattice size, especially the smaller size of the center band.

\textit{The fractal filling positions of gaps.}
Now that the organizing rules of the electron fillings is summarized, the next step is to uncover the transformation of the self-similar spectrum in terms of the filling positions of gaps.
Since the filling positions depend on the frequency $\alpha$, the self-similar relation of the fractal spectrum also depends on $\alpha$. 
Considering the golden ration frequency $\alpha=(\sqrt{5}-1)/2$, the fractal fillings could be generated by the following transformation steps:
\begin{enumerate}
    \item Take the initial spectrum as three bands separated by two gaps at fillings $\alpha,1-\alpha$, two symmetric golden ratio fillings.
    \item Transform each current band into three more bands with relative filling positions of the two gaps at $\alpha,1-\alpha$.
\end{enumerate}
To be quantitative, the gap positions could be classified into layers: 2 gaps in the first layer ($\alpha, 1-\alpha$); $6$ gaps in the second layer ($\{2\alpha\},\{3\alpha\},\{4\alpha\},\{2(1-\alpha)\},\{3(1-\alpha)\},\{4(1-\alpha)\}$); $18$ gaps in the third layers and so on. Thus, the $k$th layer will have $3^{k-1}\cdot 2$ new gaps and spectrum with $k$ layers will have $3^k-1$ gaps in total.

Note that $\{n\alpha\}=1-\{n(1-\alpha)\}$ due to the identity $\{n\alpha\}+\{-n\alpha\}=1$. If we sort the second layer, $\{3(1-\alpha)\}\approx0.1459,\{2\alpha\}\approx0.2361,\{4\alpha\}\approx0.4721, \{4(1-\alpha)\}\approx0.5279,\{2(1-\alpha)\}\approx0.7639,\{3\alpha\}\approx0.8541$, each pair locates within the previous bands $[0,(1-\alpha)],[(1-\alpha),\alpha],[\alpha,1]$. Thanks to the golden ratio relation, $\alpha^2=1-\alpha$, the second layer could be rewritten as 
\begin{equation}
\begin{split}
        \{3(1-\alpha)\}&=2-3\alpha=(1-\alpha)^2,\\
        \{2\alpha\}&=2\alpha-1=(1-\alpha)\alpha,\\
        \{4\alpha\}&=4\alpha-2=(1-\alpha)+(2\alpha-1)(1-\alpha),\\ \{4(1-\alpha)\}&=3-4\alpha=(1-\alpha)+(2\alpha-1)\alpha,\\
        \{2(1-\alpha)\}&=2-2\alpha=\alpha+(1-\alpha)^2,\\
        \{3\alpha\}&=3\alpha-1=\alpha+(1-\alpha)\alpha.
\end{split}
\end{equation}
From the above expression, we could see that the second layer of gaps happen to locate at the relative position of $1-\alpha, \alpha$ of the previous band, which corresponds to one transformation of the previous band. The above six gaps are all at the second order positions with highest order $\alpha^2$. The next two gap positions $\{5\alpha\}=5\alpha-3=0+ (1-\alpha)^2\alpha$, $\{5(1-\alpha)\}=4-5\alpha=\{3\alpha\} +(1-\alpha)^3$ are in the third layer. The gaps $\{n\alpha\},\{n(1-\alpha)\}$ belong to $k$th layer if the integer $n$ satisfies
\begin{equation}
    \frac{3^{k-1}-1}{2}<n<=\frac{3^{k}-1}{2}.
\end{equation}

This self-similarity could also be confirmed by the two transformation function of the Hofstadter butterfly, which only works at the critical point.
In the fractal butterfly\cite{PhysRevB.14.2239}, the spectrum of frequency $\alpha$ could be separated into three parts: the left (L), right (R) and the center (C) parts, where each part could be a scaled spectrum of another two frequencies $\alpha'$(L,R parts),$\beta'$(C part) based on the two functions
\begin{equation}
    \begin{split}
        \alpha'&=\frac{1}{\alpha}-[\frac{1}{\alpha}]=\frac{\sqrt{5}-1}{2}=\alpha\\
        \beta'&=(\frac{1}{\alpha}-2)^{-1}=-\frac{\sqrt{5}+3}{2}\to \frac{1-\sqrt{5}}{2}=1-\alpha,
    \end{split}
\end{equation}
where $[x]$ denotes the greatest integer less than or equal to $x$. Thus, with the golden ratio frequency, the new L, R and C parts are generated by the same/conjugate incommensurate $\alpha$ and the self-similar transformations are confirmed. 

Now that the fractal fillings for $\alpha=(\sqrt{5}-1)/2$ is revealed, can we extend the self-similar transformations to arbitrary irrational frequencies $\alpha$? From the rules of gap filling positions, the frequency should be able to transform $\{n\alpha\}$ into higher order representation $\alpha^k$, namely the irrational frequency for a fractal spectrum should be at least a solution of a polynomial equation. From the perspective of the butterfly transformations, the irrational frequency should return to the same value after several iterations of transformations. Another example for the fractal spectrum is $\alpha=\frac{\sqrt{2}-1}{2}$. We could find that self-similar pattern formed around the center of the spectrum (the gap position $2\alpha=\frac{1}{2}-(\frac{1}{2}-\alpha)^2$, or $\alpha',\beta'=\alpha$ after two transformation of the two function.) However, the frequency $\alpha=\ln(2)$ could never give such self-similar relations.

\textit{The scaling of gap sizes.}
From the above analysis, the fractal filling positions of gaps only concerns the special irrational numbers. However, the fractal spectrum only manifests itself at the critical point $\lambda=2$ of the AA model. The scaling of gap sizes with higher multiples $n$ (or higher layers $k$) for various strength $\lambda$ becomes critical. From the self-similar relations at the critical point, we could derive that the gap sizes should follows a power-law decay. From the initial spectrum, the three bands (L, C, R parts) are just scaled spectrum of the whole spectrum by two empirical factors
\begin{equation}
\begin{split}
        f_{L/R}&=\frac{E_{max}-E_{side}}{2E_{max}}=0.0729,\\
        f_C&=\frac{E_{mid}}{E_{max}}=0.1392,
\end{split}
\end{equation}
where $E_{max}=2.5975$,$E_{side}=1.8742$, $E_{mid}=0.18935$ denotes the half band width of the whole spectrum, starting energy of the side band and the half band width of the center band, respectively.
Each new layer of gaps could be generated by multiplying $f_{L/R}$ of the previous gaps for L, R parts and $f_{C}$ for C part. Since each current gap will be transformed into three new gaps, we could verify the scaling by the self-consistent relation:
\begin{equation}
\begin{split}
        E_{max}&= \sum_{k=0}^\infty (E_{side}-E_{mid})(2f_{L/R}+f_{C})^k\\
        &=\frac{E_{side}-E_{mid}}{1-(2f_{L/R}+f_{C})},
\end{split}
\end{equation}
where $(E_{side}-E_{mid})$ is the size of the first two gaps.

\begin{figure}[t!]
\begin{center}
\includegraphics[width = 0.49\textwidth]{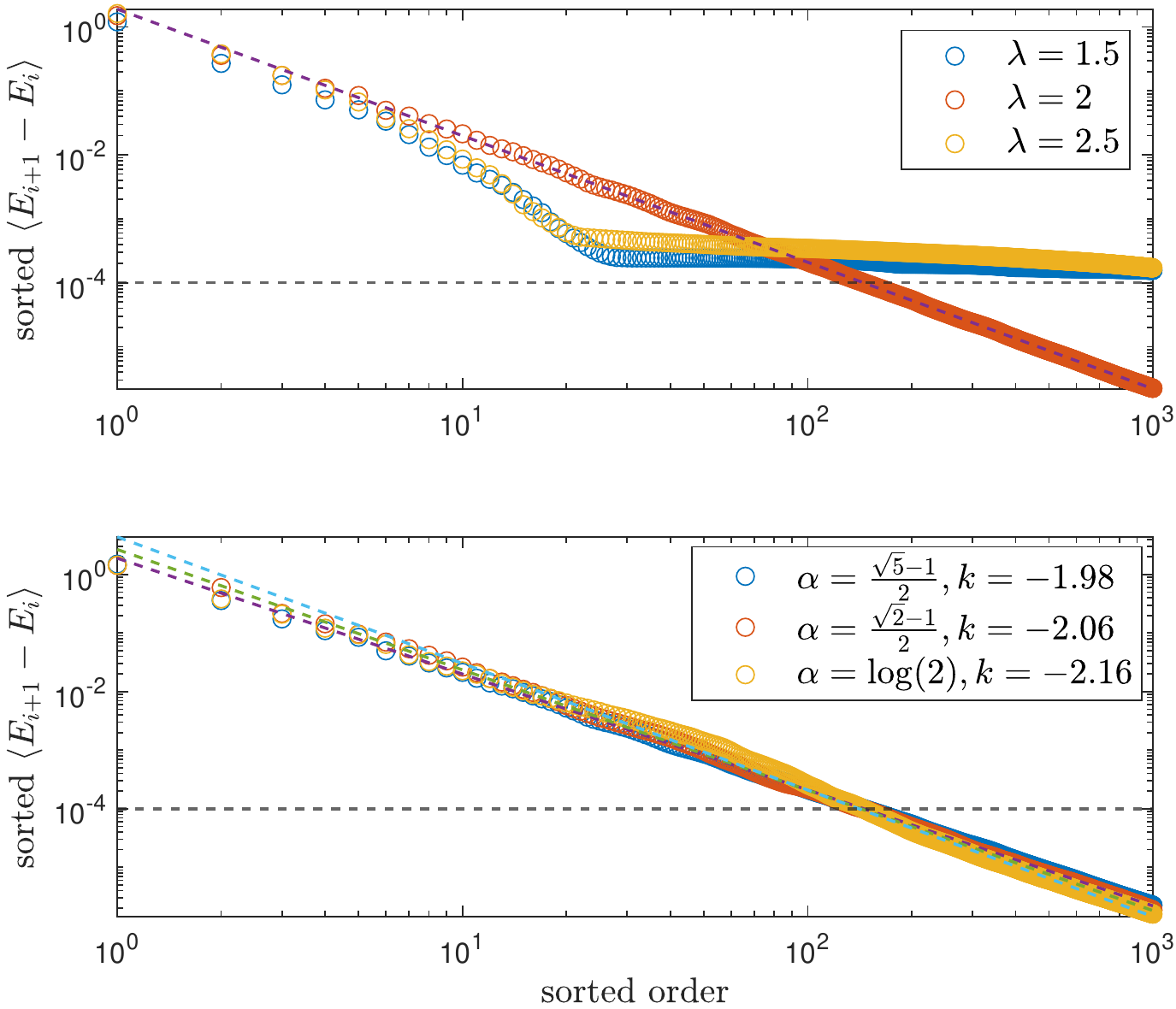}
\caption{The scaling behavior of sorted nearest state gaps of the half spectrum. Only the first $1,000$ gap sizes are shown out of total $L/2=5,000$ gaps. The gaps are averaged over $100$ random phases $\phi$. The horizontal dashed line marks $\delta=1/L$, the energy resolution scale. The dashed lines are linear fits of the sorted gaps as a function of order index. Upper panel: The gap scaling in different phases. Lower panel: The gap scaling of various frequency at the critical point $\lambda=2$. }
\label{gapsorted}
\end{center}
\end{figure}

If we sort the nearest state gaps by descending order with index $n$, the scaling of gap size are shown in Fig. \ref{gapsorted}. In both the localized and delocalized phase, the gap sizes quickly converge to the generic energy resolution $\delta=1/L$. The critical point gives a clear power-law scaling of gap sizes to a scale lower than the energy resolution, with a power law
\begin{equation}
    \delta_n\sim n^{-1.98}.
\end{equation}
Note that the gap sizes decay in an exponential fashion as a function of the layer order $k$, but the sorted order index $n$ relates to the layer order in a logarithmic way, $k=1+[\log_3(n)]$, for $\alpha=(\sqrt{5}-1)/2$. In the large $n$ (or $k$) limit, the dominating term is $(\sqrt[3]{f_{L/R}^2f_{C}})^k$, which gives
\begin{equation}
    \delta_n\sim (f_{L/R}^2f_{C})^{k/3} \sim n^{-2.19}.
\end{equation}
The deviation of the power-law comes from the finite layer $k=[\log_3(1,000)]=6$ and the finite lattice size. The power-law scaling of gaps makes it possible for the self-similar transformations among small and large scales through simple multiplication factors. However, this power-law behavior does not limited by this particular irrational frequency. It appears at different irrational frequencies, with and without self-similar structures, as shown in the lower panel of Fig. \ref{gapsorted}. In other words, the power-law scaling is characteristic at the critical point while the fractal spectrum is not universal.

In conclusion, we reveal the fractal spectrum of the AA model from the aspect of the fractal filling. When a periodic potential is applied to a lattice system, the band will be separated into multiply bands according to the periodicity of the potential, which could be mapped to a multi-atom system with translational symmetry. When the periodicity is an irrational number, the translational symmetry is broken and the system is separated into infinite mini-bands. The electron fillings of these mini-bands are just multiples of the frequency $\alpha$ and its counterpart $1-\alpha$ with a fold into the range $[0,1]$.

With specific irrational frequency, e.g. $\alpha=(\sqrt{5}-1)/2$, the filling positions of gaps are fractal: more gaps could be generated by iteratively inserting large scale structure into mini-bands. However, these fractal fillings are limited to certain types of irrational numbers, solutions of polynomial equations. 

The fractal fillings depend only on the irrational frequencies of the potential, but the fractal spectrum requires the mini-gap sizes decay in a power-law fashion which depends on the strength of the quasi-periodic potential. The critical point of the AA model happens to give rise this power-law decay of gap sizes.

This work separates the study of the fractal spectrum into two independents parts. The fractal spectrum of the AA model requires both a proper incommensurate frequency and the power-law decay of the emerging gaps. The first gives the fractal filling of the spectrum; the second makes the fractal filling manifested in the DOS. This study of the fractal spectrum shall shad lights on studies of incommensurate systems.

\textit{Acknowledgements.} The author is grateful to Jedediah H. Pixley for encouraging this project as an independent work. The author also thanks Eric Gawiser for taking this project as a course project in his astrophysics class and thanks Ghanashyam Khanal for coding neural networks to conduct density estimation on the fractal DOS. This work was supported by NSF CAREER Grant No. DMR1941569. 

\bibliography{ref.bib}

\end{document}